# How two-dimensional bending can extraordinarily stiffen thin sheets

V. Pini[†], J.J. Ruz[†], P. M. Kosaka, O. Malvar, M. Calleja and J. Tamayo[*]

Institute of Microelectronics of Madrid (IMM-CSIC), Isaac Newton 8 (PTM), Tres Cantos, 28760 Madrid, Spain.

**Curved thin sheets are ubiquitously found in nature and manmade structures. Within the framework of classical thin plate theory, the stiffness of thin sheets is independent of its bending state. This assumption, however, goes against intuition. Simple experiments with a cantilever sheet made of paper show that the cantilever stiffness largely increases with the transversal curvature. We here demonstrate by using simple geometric arguments that thin sheets subject to two-dimensional bending necessarily develop internal stresses. The coupling between the internal stresses and the bending moments can increase the stiffness of the plate by several times. We develop a theory that describes the stiffness of curved thin sheets with simple equations in terms of the longitudinal and transversal curvatures. The theory perfectly fits experimental results with a macroscopic cantilever sheet as well as numerical simulations by the finite element method. The results shed new light on plant and insect wing biomechanics and provide an easy route to engineer micro- and nanomechanical structures based on ultrathin materials with extraordinary stiffness tunability.**

### Thin plate mechanics with a paper sheet

A paper sheet is a paradigmatic example of thin elastic sheets that can bend in a host of different and complex patterns with small forces[1-5]. Whereas the role of internal stress in the three dimensional configurations of thin sheets is starting to be understood, the reverse effect has received little attention as shown here. In a simple experiment with a cantilever made of a paper sheet, we find two properties that cannot be explained with current theoretical models on plate mechanics (Fig. 1). Our paper sheet is 297 mm long, 210 mm wide and 340 μm thick. We not that the problem is scalable and the same phenomena can, for instance, be found in a microfabricated thin plate 297 μm long, 210 μm wide and 340 nm thick. The paper cantilever is obtained by fixing one of the ends of the sheet to a desk. The cantilever largely bends downwards due to gravity (Fig. 1a). If we add a small transversal curvature of 0.0035 mm$^{-1}$ with positive sign (against gravity) to the clamped cantilever end, the sheet gets almost straight (Fig. 1b). This experiment reveals the first property: *the stiffness of cantilever sheets significantly increases with the transversal curvature*. We now rotate 180 degrees the transversally curved cantilever sheet. The cantilever deflects downwards due to the gravity, more than before rotation, but substantially less than in the flat configuration (Fig. 1c). This experiment shows the second property referred to as *bending asymmetry: the upward and downward bending stiffness of a cantilever sheet is different when the cantilever is transversally curved.*

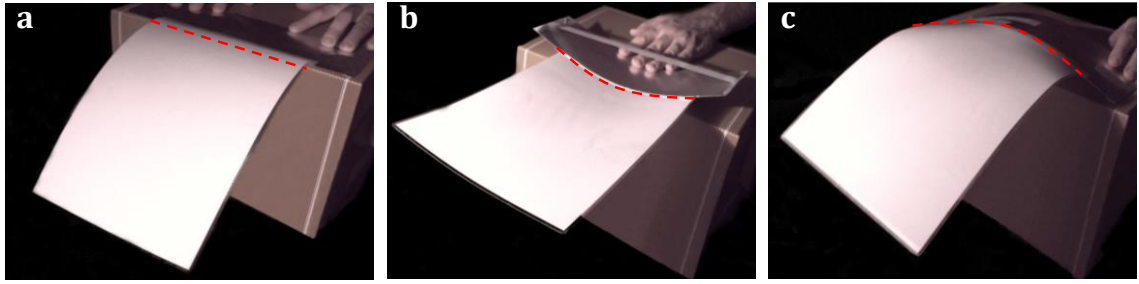

**Figure 1| Stiffening of a paper sheet induced by transversal bending. a,** Photographs of a A4 paper cantilever sheet subject to gravity force when the transversal curvature is zero (a), positive (b) and negative (c).

The elastic energy for small deflection of a thin cantilever plate of length *L*, width *b* and thickness *h* is given by $U = \frac{1}{2}\iint \left\{ D_{ijkl}\partial_{i,j}w\, \partial_{k,l}w - \frac{1}{2}\sigma_{ij}h\, \partial_i w\, \partial_j w \right\} dxdy$, where *w* is the deflection in the z direction, the index *i*, *j*, *k* and *l* range from 1 to 2 and denote the in-plane *x* and *y* directions, $D_{ijkl}$ is the flexural rigidity tensor, $\sigma_{ij}$ is the in-plane stress tensor at the mid surface[6,7]. The first term and second term are the bending energy and in-plane stress energy, respectively. In the case of longitudinal bending and longitudinal stress, bending energy scales as $\sim Eb\left(\frac{h}{L}\right)^3$, where *E* is the Young's modulus; and the stress energy scales as $\sim \sigma b\left(\frac{h}{L}\right)$. Since the stiffness of the plate is the second derivative of the potential energy with respect to the z-displacement, the stiffness has the form $k = \alpha_1 Eb\left(\frac{h}{L}\right)^3 - \alpha_2 \sigma b\left(\frac{h}{L}\right)$, where $\alpha_1$ and $\alpha_2$ are numerical constants that depend on the deflection shape. In the case of cantilevers, the second term of the stiffness is marginal as most of the in-plane stress is released through deformation of the free end[6]. *Therefore in the framework of classical plate theory, the stiffness of thin sheets with no in-plane stresses depends on the mechanical properties and dimensions of the plate and it is independent of the bending configuration.* This statement leaves the results with the paper cantilever unexplained.

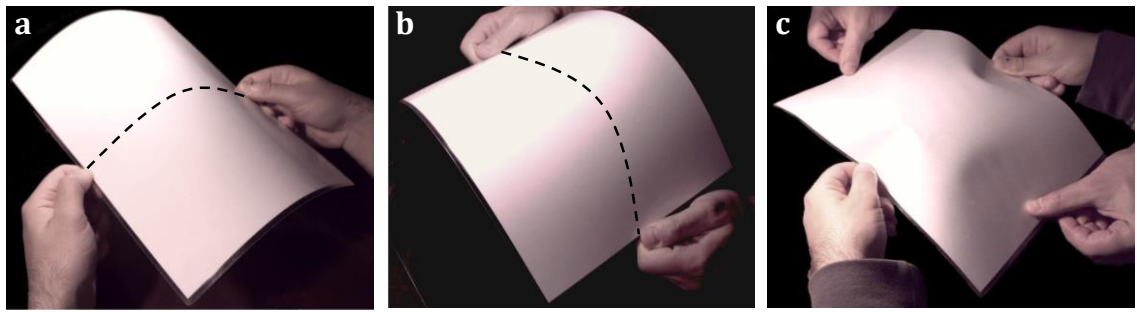

**Figure 2| Wrinkling of a paper sheet subject to biaxial bending.** Photographs of a A4 paper sheet manually subject to uniaxial bending in the transversal (a) and longitudinal (b) directions; and to biaxial bending (c).

To understand why the paper cantilever stiffens when it is transversally curved, we make a second experiment with the paper sheet. We observe that whereas the sheet can be easily bent in one direction within the elastic regime (Figs. 2a and 2b), biaxial bending is not so straightforward (Fig. 2c). Far from achieving a parabolic-like surface[8], our sheet exhibits a characteristic wrinkle pattern that impedes biaxial bending. This experiment reveals a third property: *in-plane strains are necessarily developed when the sheet is subject to biaxial bending*. We will show that this property is the origin of the stiffening of the paper cantilever when it is transversally bent. First, we demonstrate by appealing to simple geometry arguments that in-plane strains are necessarily generated when a sheet is subject to bending moments in the two plane directions. We start with a thin sheet with length, width and thickness referred to as $L$, $b$ and $h$, respectively, (Fig. 3). The structure is unrestrained at the edges and it is only subject to bending moments. The original flat structure must then transform into a structure with out-of-plane displacement $w_s(x,y)$ that obeys the following three topological constraints; the volume, contour length and contour width must remain constant. These constraints are consequence of that no in-plane stresses are applied to the sheet. For small displacements ($w_s(x,y) \ll L, b$), the constraints are mathematically expressed by applying Pythagorean theorem to infinitesimal lengths in the chosen Cartesian coordinate system (Fig. 3),

$$bhL = \int_0^{L_x} \int_{-\frac{b_y}{2}}^{\frac{b_y}{2}} h_z dx dy \tag{1a}$$

$$L = \int_0^{L_x} \left\{1 + \frac{1}{2}(\partial_x w_s)^2\right\} dx \tag{1b}$$

$$b = \int_{-\frac{b_y}{2}}^{\frac{b_y}{2}} \left\{1 + \frac{1}{2}(\partial_y w_s)^2\right\} dy \tag{1c}$$

where $L_x$ and $b_y$ are the projections of the length and the width of the deformed sheet onto the $x$- and $y$- axes, respectively; and $h_z$ is the thickness along the $z$- axis given by,

$$h_z = h\left\{1 + \frac{1}{2}(\partial_x w_s)^2 + \frac{1}{2}(\partial_y w_s)^2\right\} \tag{2}$$

Since there are more equations than unknowns, $L_x$ and $b_y$, the system of equations, Eqns. (1a)-(1c), is overdetermined and it is, in general, inconsistent. This is revealed by substituting Eqns. (1b), (1c) and (2) into Eq. (1a) which results in the paradoxical equation,

$$(L_x - L)(b_y - b) = 0 \tag{3}$$

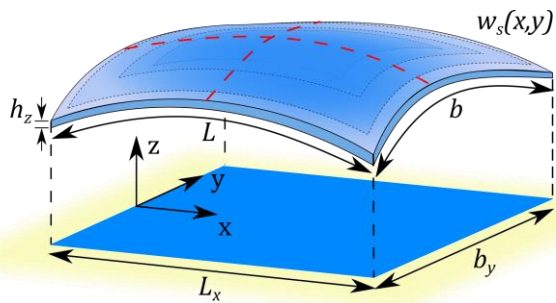

**Figure 3| Schematic of the bending of an unrestrained thin sheet.** The sheet in the flat state has length, width and thickness given by $L$, $b$ and $h$ respectively. Biaxial bending induces out-of-plane displacement (z direction) given by $w_s(x,y)$. The projections of the length and the width of the deformed sheet onto the $x$- and $y$- axes are given by $L_x$ and $b_y$, respectively. The origin of the $x$-axis is at the left free edge of the undeformed plate, and the origins of the $y$- and $z$- axis are at the middle of the undeformed plate.

Since $L_x \leq L$ and $b_y \leq b$, equation (3) can only be satisfied if the sheet is bent in one direction, as the length of the sheet in the orthogonal direction remains constant. Equation (3) is not satisfied in the case of biaxial bending, which implies that strains are necessarily generated, modifying the contour length of the sheet and thus violating Eqns. (1b) and (1c).

**Theory of the stiffness of curved thin sheets**

In thin sheets, the energy cost of in-plane straining is very high, $\sim \frac{h}{L}$, in comparison with the bending energy $\sim \left(\frac{h}{L}\right)^3$. The most obvious manifestation of this effect is that small in-plane stresses in thin plates give rise to formation of three dimensional configurations such as buckling or wrinkling-like instabilities[1,2,5]. We demonstrate the opposite phenomena; three dimensional configurations in thin plates give rise to internal strain configurations. The curvature-induced in-plane strain significantly increases the elastic energy, leading to stiffening as shown below. The resolution of the strain that minimize the elastic energy cost in thin sheets poses a formidable theoretical challenge. We circumvent the enormous difficulty of the problem by starting with the thin sheet already deformed and calculating the energy cost associated to an additional small deformation. The initial deformation is referred to as static deformation and it may be plastic or/and elastic. The associated strains may also include geometric nonlinearities[9]. The second deformation is referred to as dynamic deformation and it must be elastic and sufficiently small to be described by linear strain-displacement relations. The out-of-plane displacements of the static and dynamic deformations are respectively given by $w_s(X,Y)$ and $\Delta w(X,Y)$, where $X = \frac{x}{L}$ and $Y = \frac{y}{b}$. Our proposed functions for the in-plane $x$ and $y$ displacements associated to the dynamic deformation are given by,

$$\Delta u(X,Y,z) = f(X,Y) - \partial_X \Delta w(X,Y) \left(\frac{z - w_s(X,Y)}{L}\right) \tag{4a}$$

$$\Delta v(X,Y,z) = g(X,Y) - \partial_Y \Delta w(X,Y) \left(\frac{z - w_s(X,Y)}{\beta L}\right) \tag{4b}$$

where $\beta \equiv \frac{b}{L}$. The first summand in Eqns. (4a) and (4b) represent the in-plane displacement of the mid surface and the second summand the bending displacement. The proposed displacement field departs with respect to classical Kirchhoff-Love plate theory[7] in the following: i) the neutral plane is substituted by the curved surface, $w_s(X,Y)$ of the bent sheet and ii) the in-plane displacement of the mid surface is nonzero, albeit in-plane stresses are not exerted on the plate. The strain tensor induced by the dynamic deformation is given by,

$$\varepsilon = \begin{pmatrix} \frac{\partial_X u}{L} & \frac{1}{2}\left(\frac{\partial_Y u}{\beta L} + \frac{\partial_X v}{L}\right) & \frac{1}{2}\left(\partial_z u + \frac{\partial_X \Delta w}{L}\right) \\ \frac{1}{2}\left(\frac{\partial_Y u}{\beta L} + \frac{\partial_X v}{L}\right) & \frac{\partial_Y v}{\beta L} & \frac{1}{2}\left(\partial_z u + \frac{\partial_Y \Delta w}{\beta L}\right) \\ \frac{1}{2}\left(\partial_z u + \frac{\partial_X \Delta w}{L}\right) & \frac{1}{2}\left(\partial_z u + \frac{\partial_Y \Delta w}{\beta L}\right) & 0 \end{pmatrix} \tag{5}$$

We calculate the elastic energy of the system by using,

$$U = \frac{1}{2} L_x b_y C_{ijkl} \int_0^{\frac{L_x}{L}} dX \int_{-\frac{1}{2}\frac{b_y}{b}}^{\frac{1}{2}\frac{b_y}{b}} dY \int_{w_s(X,Y) - \frac{1}{2}h_z}^{w_s(X,Y) + \frac{1}{2}h_z} \varepsilon_{ij} \varepsilon_{kl} dz \tag{6}$$

where $C_{ijkl}$ is the elastic tensor and Einstein summation convention is used. The unknown displacement functions, $f$ and $g$ are obtained by solving the Euler-Lagrange equations corresponding to the elastic energy of the system. To find analytical solutions to the problem, the following assumptions must be adopted: i) the deformation of the plate is symmetric about the y-axis of the thin plate and ii) the y-curvatures of the static and dynamic z displacements are independent of the *y*-coordinate. We refer the *x*- and *y*- directions to the longitudinal and transversal directions of the plate. For the sake of simplicity, we also assume that the plate is an isotropic material. The analytic solution of the problem is detailed in the Supplementary Materials (Sect. S1). We here provide the resulting potential energy that takes the form, $U = U_b + U_s$, where $U_b$ is the bending energy and $U_s$ is the in-plane straining energy. The bending energy is given by,

$$U_b = \frac{EL^5 \beta \eta^3}{24(1-\nu^2)} \int_0^1 \{\Delta \kappa_y(X)^2 + 2\nu \Delta \kappa_y(X) \Delta \kappa_x(X) + \Delta \kappa_x(X)^2\} dX \qquad (7)$$

where $\eta = \frac{h}{L}$, $E$ is the Young modulus, and $\nu$ is the Poisson's ratio; $\Delta \kappa_x$ is the dynamic *x*-curvature at the longitudinal axis of the plate and $\Delta \kappa_y$ is the dynamic *y*- curvature. Importantly, the bending energy cost due to the dynamic deformation is independent of the previous static deformation of the plate.

The straining energy reads as,

$$U_s = \frac{EL^7 \beta^5 \eta}{1440} \int_0^1 \big(\kappa_x(X)\Delta \kappa_y(X) + \kappa_y(X)\Delta \kappa_x(X)\big)^2 dX \qquad (8)$$

where $\kappa_x(X)$ is the static *x*-curvature at the longitudinal axis of the plate and $\kappa_y(X)$ is the static *y*- curvature. Interestingly, the straining energy depends on the orthogonal coupling between the static and dynamic curvatures. The straining energy is zero when the sheet is only bent in one direction (e.g. $\Delta \kappa_y = \kappa_y = 0$). This is in consistency with the topological constraints described above; uniaxial bending is compatible with zero straining of the neutral surface. However, in-plane strains necessarily comes up when the sheet, for instance, is initially bent in the transversal direction and then it is longitudinally bent. The straining energy arises from in-plane stresses that act as a pulling force against the longitudinal bending.

We analyze the practical case, in which i) the initial static curvatures are constants and ii) the curved sheet is subject to longitudinal bending moment, so $\Delta \kappa_y(X) \approx -\nu \Delta \kappa_x(X)$. The relative change of stiffness in this case is independent on how the load is distributed along the plate in the dynamic deformation step,

$$\frac{\Delta k}{k} = \frac{U_s}{U_b} = \frac{L^2}{60} \frac{\beta^4}{\eta^2} \big(\kappa_y - \nu \kappa_x\big)^2 \qquad (9)$$

Equation (9) was compared with numerical results by the finite element method (FEM), showing an error below 10% for free plates with $\beta < 0.4$ (Supplementary Sect. S2). A glance at equation (9) allows direct understanding of the basic mechanisms of plate stiffening due to curvature. The increase of the stiffness scales like $\sim \frac{\beta^4}{\eta^2}$. Thus, in the limit of very thin plates $\eta \ll 1$, the stiffness may be dominated by the initial curvatures of the plate. The quartic dependence on $\beta$ indicates that this is a purely plate effect; the stiffening goes to zero in the limit of narrow beams ($\beta \ll 1$). The effect of longitudinal curvature goes as $\sim \nu^2$, being significantly smaller than the effect of transversal curvature. Transversal curving seems to be the most economical way to stiffen a thin plate structure. This may be the case of plant leaves. It is well-known the leaf-shapes have been optimized during evolution to absorb sufficient light and facilitate gas exchange[10]. From this point of view, leaves must be as wide, flat and thin as possible. However, leaf shape must also be optimized from the mechanical point of view; the leaf must support its own weight and external dynamic forces such as the pressure of winds[11]. Leaf curvature is genetically controlled and it can be modified at wish in some species by gene mutation[12]. The large influence of curvature on the stiffness of thin and wide sheets suggests that leaf curvature has been targeted by evolution for optimization of their biological and structural functions. The effect of bending asymmetry observed with the paper sheet (Fig. 1) can be explained by the coupling term in Eq. (9), $\sim -\nu \kappa_x \kappa_y$. The bending stiffness of a thin plate with *x*- and- *y*- curvatures of opposite sign is higher than in the case of *x*- and *y*-curvatures with the same sign. Bending asymmetry plays a key role in the flight of insects. Detailed FEM simulations of insect wings have shown that chordwise camber of the insect wings induces significant wing stiffening and the coupling between chordwise and spanwise camber is the origin of the dorsal-ventral bending asymmetry[13]. Here, we provide simple analytical equations that enable an intuitive understanding of this phenomenon.

Equation (9) was deduced for plates with free edges. However, it is common in nature and engineering to find these structures with one of the edges clamped. In these cases, the transversal curvature is zero at the clamping end and it exponentially increases in the $x$ direction with a characteristic length given by the cantilever's width, tending to reach the asymptotic value $(\kappa_y)$[14]. We accordingly modify equation (9) to include the clamping effect,

$$\frac{\Delta k}{k} = \frac{L^2}{60} \frac{\beta^4}{\eta^2} e^{-c\beta} \left( \kappa_y - \nu \kappa_x \right)^2 \tag{10}$$

The constant $c$ depends on the shape of the dynamic deformation and must be computed by fitting Eq. (10) to FEM simulations. For the case of flexural bending in the fundamental vibration mode, $c \approx 3.095$. The error of Eq. (10) is below 5% when compared with FEM simulations for $\beta < 0.4$ (Supplementary Sect. S3).

**Comparison of the theory with a experiment based on a macroscopic cantilever plate**
We performed an easy experiment with a homebuilt macroscopic aluminum cantilever plate to validate our theoretical predictions (Fig. 4a). A step force was manually applied to the cantilever free end and the transient vibration was recorded by a smartphone camera (Fig. 4b) (Suppl. Sect. S4). The resonant frequency of a plate is related to the spring constant, $k$, by the harmonic oscillator equation $\omega_o = \sqrt{k/m}$ where $m$ is the mass of the plate. Figure 4c summarizes the values of the resonance frequency for different curvature configurations. The flat cantilever plate at vertical orientation shows the lowest resonance frequency of about $4.4\ Hz$. When the cantilever is horizontally oriented, gravity induces longitudinal bending and the resonance frequency slightly increases to *4.9 Hz*. The aluminum sheet was then transversally bent with very small curvature ($\kappa_y$=1.15 m$^{-1}$, see inset in Fig. 4a) and then clamped. At the vertical orientation, the resonance frequency of the transversally curved cantilever largely increased to *8.4 Hz* that is equivalent to an increase of the stiffness of *3.6* times. This proves the fundamental role of the transversal curvature in the stiffness of thin plates. To observe the bending asymmetry effect, the transversally curved cantilever was horizontally oriented with the transversal curvature inwards and outwards gravity. In consistency with our theory, in the first case, the resonance frequency decreases *0.7 Hz*, with respect to the transversally bent cantilever at the vertical position, whereas, in the second case the resonance frequency increases *0.3 Hz*. In order to quantitatively compare these results with the theory, we must extend our theoretical model to the case in which the static curvatures are not uniform as it is the case of gravity-induced bending (Suppl. Sect. S4). The comparison between theory and experiment shows a very good agreement (Fig. 4d).

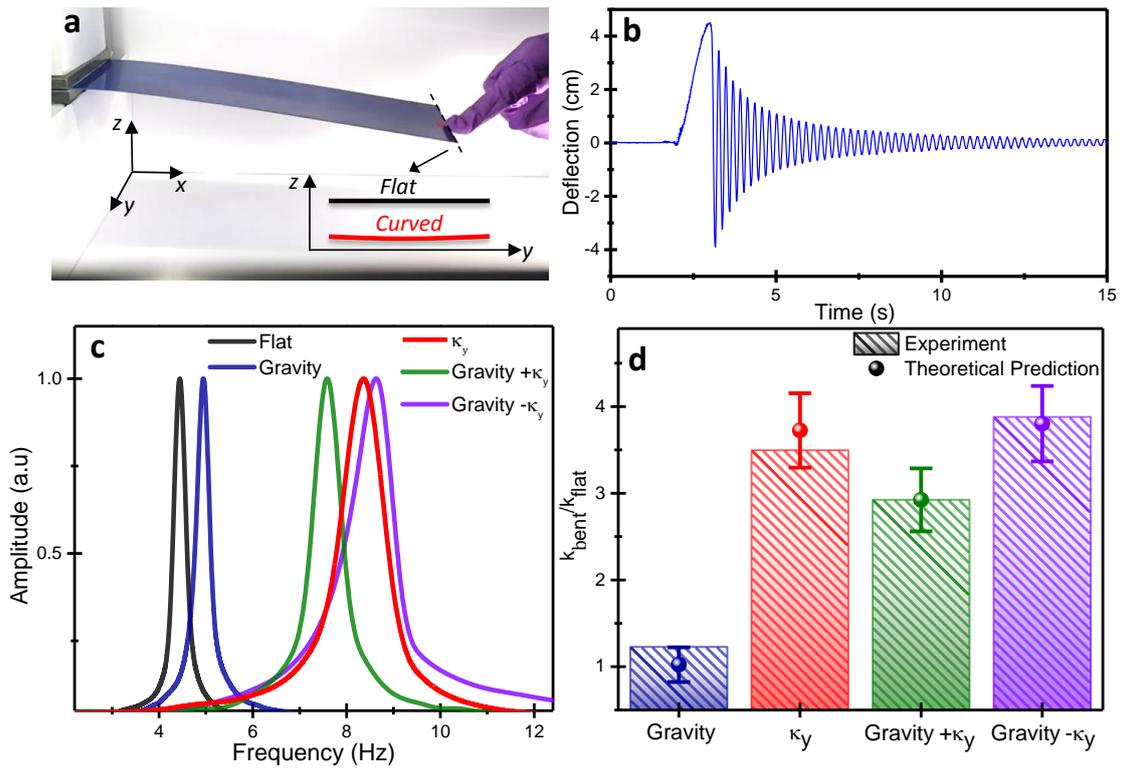

**Figure 4| Experiment with a home-made cantilever plate. a,** Photograph of the macroscopic cantilever plate. The dimensions of the cantilever are 29 ×10 ×0.0645 cm$^3$. The cantilever was set into oscillation by applying a step force at the free end by one of our fingers. The inset shows the scaled profile of the free end of the cantilever in *yz* plane in the flat and curved configurations. **b,** Transient oscillation of the microcantilever plate. **c,** Frequency response of the cantilever at the different curvature configurations. **Flat** denotes the flat cantilever vertically oriented with respect to gravity; **Gravity** label the flat cantilever in horizontal position; $\kappa_y$ denotes the cantilever with transversal curvature (inset of **a**) at the vertical position; **Gravity**+$\kappa_y$ and **Gravity**−$\kappa_y$ denotes the transversally bent cantilever at the horizontal position with curvature inward and outward gravity, respectively. **d,** Comparison between theory and experiment of the ratio between the stiffness of the bent cantilever and the flat cantilever at vertical position.

**Conclusions**

Here, we demonstrate that small amounts of two-dimensional bending of thin plates lead to the development of in-plane stresses, and thus bending is not in general so energetically cheap as believed hitherto. Biaxial curving of thin sheets is perhaps the most efficient way to reinforce the mechanical stability of thin sheets. This ancient mechanism found in nature and in man-made structures has remained unexplained in physics. The theoretical model and simple equations provided here solve this long-standing problem shedding light in our understanding of thin sheet mechanics and bringing new and useful mathematical tools to a wide variety of fields such as natural science[15], mechanical engineering and micro/nanotechnology[16-20].

**Acknowledgements**

We thank to David Fuster for his help in the cantilever experiment. This work was supported by the Spanish Science Ministry (MINECO) through project MAT2012-36197; and by European Research Council through NANOFORCELLS project (ERC-StG-2011-278860).

**Contributions**

V.P. J.J.R., M.C. and J.T. conceived and designed the work. V.P. J.J.R. and J.T. developed the theoretical model. V.P. J.J.R. carried out the FEM simulations. O.M. V.P., J.J.R. and P.K. carried out the experiments. J.T. wrote the manuscript with inputs from all authors. All the authors analysed the data, discussed the results and commented on the manuscript.

**Competing financial interests**

The authors declare no competing financial interests.

**Corresponding author**

Correspondence to: jtamayo@imm.cnm.csic.es

## -SUPPLEMENTARY MATERIALS-

# How two-dimensional bending can extraordinarily stiffen thin sheets

V. Pini[†], J.J. Ruz[†], P. M. Kosaka, O. Malvar, M. Calleja and Javier Tamayo[*]

*IMM-Instituto de Microelectrónica de Madrid (CNM-CSIC), Isaac Newton 8, PTM, E-28760 Tres Cantos, Madrid, Spain*

*Corresponding author: jtamayo@imm.cnm.csic.es

## S1. Stiffness and resonance frequency of a deformed plate.

We start with a plate with length $L$, width $b$ and thickness $h$. The plate is isotropic, homogenous and uniform. The Young's modulus and Poisson's ratio are given by $E$ and $\nu$, respectively. A schematic of the plate and the coordinate system is shown in Fig. 2 of the main text. The length, width and thickness of the undeformed plate are oriented along $x$-, $y$- and $z$-axes, respectively. The origin of the $x$-axis is at the left free edge of the undeformed plate, and the origins of the $y$- and $z$- axis are at the middle of the undeformed plate. We define the normalized coordinates $X = \frac{x}{L}$ and $Y = \frac{y}{b}$. The deformed plate displays an out-of-plane displacement $w_s(X,Y) = \Omega L\, n(X,Y)$, where $\Omega$ is the maximum static displacement normalized to the plate length and $n(X,Y)$ is the dimensionless function that describes the deformation shape. This initial deformation is referred to as static deformation. We then induce an arbitrarily small deformation, referred to as dynamic deformation, which is characterized by an out of plane displacement $\Delta w(X,Y)$. The relevant dimensionless parameters of the problem are $\beta \equiv \frac{b}{L}$ and $\eta \equiv \frac{h}{L}$. The proposed $x$ and $y$ in-plane displacements induced by the dynamic deformation are given by,

$$u(X,Y,z) = \Omega\, u_0(X,Y) - \frac{\partial \Delta w(X,Y)}{\partial X}\left(\frac{z - \Omega L n(X,Y)}{L}\right) \tag{S1}$$

$$v(X,Y,z) = \Omega\, v_0(X,Y) - \frac{1}{\beta}\frac{\partial \Delta w(X,Y)}{\partial Y}\left(\frac{z - \Omega L n(X,Y)}{L}\right) \tag{S2}$$

The Euler-Bernoulli beam bending relations are recovered when the plate is undeformed ($\Omega = 0$). The functions $f(X,Y)$ and $g(X,Y)$ used in the main text are $f(X,Y) = \Omega\, u_0(X,Y)$ and $g(X,Y) = \Omega\, v_0(X,Y)$.

The strains induced by the dynamic deformation are given by,

$$\epsilon_{xx} = \frac{\partial_X u}{L} \qquad \epsilon_{yy} = \frac{\partial_Y v}{\beta L} \qquad \epsilon_{xy} = \frac{1}{2}\left(\frac{\partial_Y u}{\beta L} + \frac{\partial_X v}{L}\right)$$

$$\epsilon_{xz} = \frac{1}{2}\left(\partial_z u + \frac{\partial_X \Delta w}{L}\right) \qquad \epsilon_{yz} = \frac{1}{2}\left(\partial_z u + \frac{\partial_Y \Delta w}{\beta L}\right) \tag{S3}$$

The stresses are given by[1],

$$\sigma_{xx} = \frac{E}{1-\nu^2}(\epsilon_{xx} + \nu \epsilon_{yy}) \qquad \sigma_{yy} = \frac{E}{1-\nu^2}(\epsilon_{yy} + \nu \epsilon_{xx}) \qquad \sigma_{zz} = 0$$

$$\sigma_{xy} = \frac{E}{1+\nu}\epsilon_{xy} \qquad \sigma_{xz} = \frac{E}{1+\nu}\epsilon_{xz} \qquad \sigma_{yz} = \frac{E}{1+\nu}\epsilon_{yz} \tag{S4}$$

The potential energy is given by,

$$U = \frac{1}{2}L_x b_y \int_0^{\frac{L_x}{L}} dX \int_{-\frac{1}{2}\frac{b_y}{b}}^{\frac{1}{2}\frac{b_y}{b}} dY \int_{w_s(X,Y)-\frac{1}{2}h_z}^{w_s(X,Y)+\frac{1}{2}h_z} (\sigma_{xx}\epsilon_{xx} + \sigma_{yy}\epsilon_{yy} + 2\sigma_{xy}\epsilon_{xy} + 2\sigma_{xz}\epsilon_{xz} + 2\sigma_{xz}\epsilon_{xz}) dz \quad (S5)$$

The limits of integration in Eq. (S5) are given by projections of the length, width and thickness of the deformed plate onto the x-, y-, and z- axes, respectively (see Fig. 3 of the main text). In the limit of small deflections, the projected dimensions are given by,

$$L_x \cong L\left(1 - \frac{1}{2}\Omega^2 \int_0^1 \partial_X n^2 dX\right)$$

$$b_y \cong \beta L\left(1 - \frac{1}{2}\left(\frac{\Omega}{\beta}\right)^2 \int_{-\frac{1}{2}}^{\frac{1}{2}} \partial_Y n^2 dY\right)$$

$$h_z = h\left(1 + \frac{1}{2}\Omega^2 \left(\partial_X n^2 + \frac{\partial_Y n^2}{\beta^2}\right)\right) \quad (S6)$$

We calculate the Euler-Lagrange differential equations obeyed by the unknown functions $u_0$ and $v_0$ for the potential energy density integrated along the projected thickness of the plate,

$$\partial_{XX}u_0 + \partial_X(\partial_X n \, \partial_X \Delta w) + \frac{1+\nu}{2\beta}\partial_{XY}v_0 + \frac{1}{2\beta^2}\{(1-\nu)(\partial_{YY}u_0 + \partial_X n \, \partial_{YY}\Delta w + \partial_{YY}n \, \partial_X \Delta w) + (1+\nu)\partial_X(\partial_Y n \, \partial_Y \Delta w)\} = 0$$

$$\partial_{XX}v_0 + \frac{2}{\beta^3(1-\nu)}\{\partial_Y(\partial_Y n \, \partial_Y \Delta w) + \beta \partial_{YY}v_0 + \beta^2[(1+\nu)\partial_Y(\partial_X n \, \partial_X \Delta w) + \partial_{XY}u_0 + (1-\nu)(\partial_{XX}n \, \partial_Y \Delta w + \partial_Y n \, \partial_{XX}\Delta w)]\} = 0 \quad (S7)$$

Functions $u_0(X,Y)$ and $v_0(X,Y)$ are involved in coupled differential equations that cannot be analytically solved. To obtain an analytical solution to the problem we expand $u_0$, $\Delta w$ and $n$ as power series in $Y$ up to the second-order, and the function $v_0$ up to the third-order term,

$$u_0(X,Y) = u_{00}(X) + \frac{1}{2}u_{02}(X)(\beta Y)^2$$

$$v_0(X,Y) = \left(v_{00}(X) + \frac{1}{2}v_{02}(X)(\beta Y)^2\right)\beta Y$$

$$n(X,Y) = n_0(X) + \frac{1}{2}n_2(X)(\beta Y)^2$$

$$\Delta w(X,Y) = \Delta w_0(X) + \frac{1}{2}\Delta w_2(X)(\beta Y)^2 \quad (S8)$$

Notice that we assume in Eqns. (S8) that the plate deformation is symmetric in the $y$- axis. The Euler-Lagrange equations provide four coupled differential equations for the unknown functions $u_{00}(X)$, $v_{00}(X)$, $u_{02}(X)$ and $v_{02}(X)$. An analytical solution can be obtained by neglecting terms $\sim O(\beta^4)$,

$$u_{00}(X) = -\int_0^X \left\{n_0'(P)\Delta w_0'(P) + \frac{\beta^2}{24}\left(n_2(P)\Delta w_0''(P) + n_0''(P)\Delta w_2(P)\right)\right\} dP \quad (S9)$$

$$u_{02}(X) = -n_2' \Delta w_0' - n_0' \Delta w_2'' - \frac{\beta^2}{120}\{2\nu(2+\nu)(n_0''\Delta w_2' + n_2'\Delta w_0'') - \Delta w_2(3n_2' + 2(2+\nu)n_0''') - n_2(3\Delta w_2' + 2(2+\nu)\Delta w_0''')\} \tag{S10}$$

$$v_{00}(X) = -\frac{\nu\beta^2}{24}(n_2 \Delta w_0'' + n_0'' \Delta w_2) \tag{S11}$$

$$v_{02}(X) = \frac{1}{3}\left(\nu\, n_0''\Delta w_2 - n_2(\nu\Delta w_0'' - 2\Delta w_2)\right) + \frac{\beta^2}{10080}\{(56 + 229\nu - 117\nu^2)(n_2\Delta w_0'''' + n_0''\Delta w_2'' + n_0''''\Delta w_2 + 2n_0'''\Delta w_2' + 2n_2'\Delta w_0''') + 3(39 - 11\nu)n_2''\Delta w_2 + 39(6 - 19\nu)n_2'\Delta w_2'\} \tag{S12}$$

We substitute Eqns. (S9)-(S12) into the potential energy of the plate Eq. (S5). The resulting potential energy is given by,

$$U = \frac{1}{24}EL\beta\eta \int_0^1 \left\{\frac{\eta^2}{1-\nu^2}\left(\Delta w_2^2 + \Delta w_0''^2 + 2\nu\Delta w_2 \Delta w_0''\right) + \frac{1}{60}\Omega^2\beta^4(\Delta w_2 n_0'' + n_2\Delta w_0'')^2\right\}dX + O(\beta\eta^3\Omega^2) + O(\beta^6\eta\,\Omega^2) \tag{S13}$$

The first summand corresponds to the bending energy ($U_b$) and the second to the straining energy ($U_b$). Higher order terms in the bending energy $\sim O(\beta\eta^3\Omega^2)$ and the straining energy $\sim O(\beta^6\eta\,\Omega^2)$ are complex expressions that can be neglected for plates with $\beta < 0.4$. The bending and straining energy can be written in terms of the curvatures of the out-of-plane displacements,

$$U_b = \frac{EL^5\beta\eta^3}{24(1-\nu^2)}\int_0^1\{\Delta\kappa_y(X)^2 + 2\nu\Delta\kappa_y(X)\Delta\kappa_x(X) + \Delta\kappa_x(X)^2\}dX \tag{S14}$$

$$U_s = \frac{EL^7\beta^5\eta}{1440}\int_0^1\left(\kappa_x(X)\Delta\kappa_y(X) + \kappa_y(X)\Delta\kappa_x(X)\right)^2 dX \tag{S15}$$

where $\kappa_x$ and $\Delta\kappa_x$ are the static and dynamic $x$- curvatures at the longitudinal axis of the plate ($Y = 0$); and $\kappa_y$ and $\Delta\kappa_y$ are static and dynamic $y$- curvatures. The relative change of stiffness $\Delta k/k$ and resonance frequency $\Delta f/f$ can be finally evaluated with the following equation,

$$\frac{\Delta k}{k} = 2\frac{\Delta f}{f} = \frac{U_s}{U_b} = \frac{1-\nu^2}{60}\frac{L^2\beta^4}{\eta^2}\frac{\int_0^1(\kappa_x\Delta\kappa_y + \kappa_y\Delta\kappa_x)^2 dX}{\int_0^1\{\Delta\kappa_y^2 + 2\nu\Delta\kappa_y\Delta\kappa_x + \Delta\kappa_x^2\}dX} \tag{S16}$$

We analyze the case, in which i) the static curvatures are constants and ii) the sheet is subject to longitudinal bending moment, $\Delta\kappa_y(X) \approx -\nu\Delta\kappa_x(X)$. The relative change of stiffness in this case is independent on how the load is distributed along the plate and reads as,

$$\frac{\Delta k}{k} = \frac{U_s}{U_b} = \frac{L^2}{60}\frac{\beta^4}{\eta^2}\left(\kappa_y - \nu\kappa_x\right)^2 \tag{S17}$$

## S2. FEM validation of Free Plate Theory

We compare Eq. (S17) with the results of thorough numerical analysis by the finite element method (FEM) using the commercial software COMSOL Multiphysics. The following boundary conditions were applied:

i) at the plane $Y = 0$, we impose zero-displacement along the $Y$ axis, i.e. $v(X, 0) = 0$
ii) at the plane $X = 0$, we impose zero-displacement along the $X$ axis, i.e. $u(0, Y) = 0$

*iii)* at the point $(0, 0, 0)$, $w = 0$.

These three boundary conditions restrict the translation and rotation of the plate, but enable free deformation with no clamping restrictions. The parameters swept in the simulations are:

i) The curvatures $\kappa_x$ and $\kappa_y$ were varied from $-4\frac{h}{L^2}$ to $+4\frac{h}{L^2}$.
ii) $\eta = h/L$ was varied from 50 to 300.
iii) $\beta = b/L$ was varied from 0.03 to 0.4.
iv) Poisson's ratio $\nu$ was varied from 0 to 0.45.

The total amount of numerical simulations is $\sim 10^5$. For each simulation we applied free tetrahedral meshing and a convergence study was performed by refining the mesh element size until the relative error in the plate eigenfrequency is below $10^{-7}$. The average number of degrees of freedom in the simulations was $\sim 10^6$. The numerical simulations were carried out following two sequential steps: i) the plate static deformation is calculated by applying anisotropic differential surface stress in order to independently control the longitudinal and transversal curvatures of the plate[2]; ii) the statically deformed structure is imported and a eigenfrequency analysis is performed. Notice that the plate deformation in a vibration mode fulfils the relation $\Delta\kappa_y(X) \approx -\nu\Delta\kappa_x(X)$ assumed in the derivation of Eq. (S17). The relation between the relative resonance frequency shift and the relative stiffness change is given by $\frac{\Delta k}{k} = 2\frac{\Delta f}{f}$. The numerical analysis shows that the accuracy of Equation (S17) in the main text is always below 10% for free plates in the range of the swept parameters. Fig. S1 plots the comparison between our theory and the FEM simulations of the dimensionless coefficient

$\zeta \equiv \frac{\frac{\Delta f}{f}}{(\kappa L)^2}$ that gives the ratio between the fractional frequency change and the normalized curvature. The comparison is performed for several values of $\beta$, $\eta$ and $\nu$; and for the case $\kappa_x = \kappa_y = \kappa$. Our theory predicts that $\zeta = \frac{(1-\nu)^2 L^2}{120}\frac{\beta^4}{\eta^2}$.

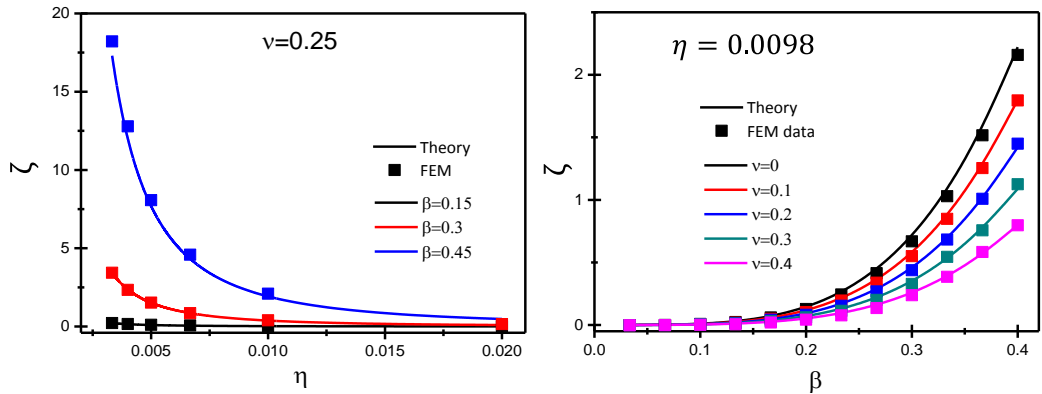

**Figure S1.** Comparison between the theory (lines) and the FEM data (symbols) of the coefficient $\zeta$ for a free plate in the case $\kappa_x = \kappa_y = \kappa$. The values of $\beta$, $\eta$ and $\nu$ are specified in the graphs.

## S3. Numerical Analysis of Cantilever Plates

The clamping effect in the case of cantilever plates is simulated by restricting the displacement in the three directions at the plane $X = 0$. The same procedure used for the free plate (Sect. S2) was followed. In this case, the FEM data was fitted to,

$$\frac{\Delta k}{k} = \frac{L^2}{60} \frac{\beta^4}{\eta^2} e^{-c\beta} \left(\kappa_y - \nu \kappa_x\right)^2 \tag{S18}$$

Numerical fittings, performed on about 12k simulations with a coefficient $R^2$ always higher than 0.999, calculate a constant value of $c = 3.095 \pm 0.001$. Equation (S18) matches the numerical data with an error below 5%.

Fig. S2 plots the comparison between our theory and the FEM simulations of the dimensionless coefficient $\zeta$ defined in Sect. S2 for several values of $\beta$, $\eta$ and $\nu$; and for the case $\kappa_x = \kappa_y = \kappa$.

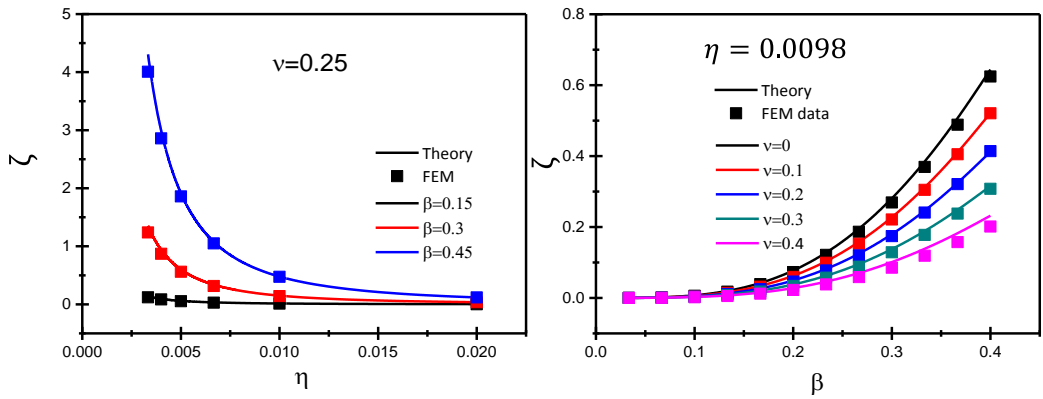

**Figure S2.** Comparison between the theory (lines) and the FEM data (symbols) of the coefficient $\zeta$ for a cantilever plate in the case $\kappa_x = \kappa_y = \kappa$. The values of $\beta$, $\eta$ and $\nu$ are specified in the graphs.

## S4. Macro Experiment: Stiffness Measurement of a Deformed Aluminum Plate

A plate was fabricated in aluminum with dimensions $29 \times 10 \times 0.0645 \; cm^3$. A permanent transversal curvature was achieved pressing the plate against a cylindrical metallic piece. Immediately after, the cantilever is clamped at one of the ends. The transversal curvature was determined by a caliper at the free end, $\kappa_y = 1.15 \pm 0.05 \; m^{-1}$. For each case the plate was anchored by one of its edges. An out-of-plane force was manually applied to the free end and then released in order to make it vibrate. The transient oscillation of the plate was imaged by a smartphone camera at a rate of 240 frames per second (IPhone 6s, Apple Inc). A region of the free end was labelled with a red adhesive. The plate oscillation is obtained by tracking the position of the red label in the obtained movie by using MATLAB software. The resonance frequency curve is obtained by calculating the fast Fourier transform of the transient oscillation. The resonance frequency is calculated by fitting the resonance peak to the damped harmonic oscillator equation. The initial longitudinal curvature is obtained by orienting the plate parallel and orthogonal with respect to the gravity force. In the case of the orthogonal orientation, the cantilever plate is subject to gravity, and the longitudinal curvature is given by $\kappa_x(X) \approx \frac{4w_L}{L^2}(X-1)^2$ where $w_L$ is the deflection of the free end[1]. The deflections in our experiments were $w_L = 2.0 \pm 0.05 \, cm$ in the case of zero transversal curvature and $w_L = 1.3 \pm 0.1 \, cm$ and $w_L = 0.1 \pm 0.05 \, cm$ for the cases transversal curvature downward and upward with respect to gravity, respectively. The difference in deflection between the downward and upward configurations comes from the bending asymmetric effect discussed in the main text. The curvatures of the dynamic deformation are given by $\Delta \kappa_x = \psi''(X)$ and $\Delta \kappa_y = -\nu \, \psi''(X)$ where $\psi(X)$ is the eigenmode shape of the fundamental mode obtained by the Euler-Bernoulli beam theory[3],

$$\psi(X) = \text{Cos}(\alpha X) - \text{Cosh}(\alpha X) + C\big(\text{Sinh}(\alpha X) - \text{Sin}(\alpha X)\big) \tag{S19}$$

where $C = 0.7340906$ and $\alpha = 1.8751$. Eq. (S16) then transforms into

$$\frac{\Delta f}{f} = \frac{L^2}{120} \frac{\beta^4}{\eta^2} e^{-c\beta} \frac{\int_0^1 \big(\kappa_y - \nu \kappa_x(X)\big)^2 \psi''(X)^2 dX}{\int_0^1 \psi''(X)^2 dX} \tag{S20}$$

The exponential term accounts for the clamping restriction. The two main sources of error of the theoretical predictions in Fig. 3d of the main text comes from the $\kappa_y$ uncertainty and from the plate thickness uncertainty, $h = 645 \pm 20 \, \mu m$.